\documentclass[12pt]{iopart}

\usepackage{harvard}
\usepackage{epsfig}
\usepackage{graphicx}
\newcommand{\cn}{\citeasnoun}
\newcommand{\np}{\newpage}
\renewcommand{\i}{\it}

\newcommand{\ba}{\begin{eqnarray}}
\newcommand{\ea}{\end{eqnarray}}
\newcommand{\be}{\begin{equation}}
\newcommand{\ee}{\end{equation}}
\usepackage{bm,amssymb}

\newcommand{\Wcm}[2]{
$\rm {#1}\times10^{{#2}}~W/cm^2$}
\newcommand{\bp}{\begin{minipage}}
\newcommand{\ep}{\end{minipage}}
\newcommand{\hs}{\hspace*}
\newcommand{\vs}{\vspace*}

\newcommand{\w}{\omega}
\newcommand{\la}{\langle}
\newcommand{\ra}{\rangle}
\renewcommand{\r}{\bm r}
\renewcommand{\k}{\bm k}
\newcommand{\A}{\bm A}
  \newcommand{\E}{{\bm E}}
  
\newcommand{\R}{{\bm R}}

\renewcommand{\k}{{\bm k}}
\renewcommand{\H}{H$_2$~}

\begin{document}

\title[The attoclock and the tunnelling time debate]
{The attoclock and the tunnelling time debate}

\author{ Anatoli S.\ Kheifets}

\address{
	Theoretical Physics, Research School of Physics, 
	The Australian National University, Canberra, Australia
	}
\ead{a.kheifets@anu.edu.au}
\vspace{10pt}
\begin{indented}
\item[]9 August -- \today
\end{indented}

\begin{abstract}

Attosecond angular streaking, also known as the ``attoclock'', employs
a short elliptically polarized laser pulse to tunnel ionize an
electron from an atom or a molecule and to put a time stamp on this
process by deflecting the photoelectron in the angular spatial
direction.  This deflection can be used to evaluate the time the
tunneling electron spends under the classically inaccessible barrier
and to determine whether this time is finite. In this review, we examine
the latest experimental and theoretical findings and present a
comprehensive set of evidence supporting the zero tunneling time
scenario. 

\end{abstract}

%
%
%
%

\tableofcontents

\np
\section{Introduction}
\markboth{Introduction}{Introduction}

Time, one of the most elusive concepts in quantum mechanics, has never
been under greater scrutiny since a recent development and application
of ultrafast pulsed laser techniques. With the temporal resolution
down to only a few attoseconds (1~as = $10^{-18}$~s), ultrafast electron
dynamics in atoms and molecules can now be probed on its native time
scale \cite{RevModPhys.81.163}.  This unprecedented experimental
capability allows one to test the most fundamental concepts at the
heart of quantum mechanics. One such concept is the tunneling time,
i.e. the time which a tunneling particle spends under the barrier in a
classically inaccessible region.

Experimental access to  tunneling time has been opened following
the pioneering ``attoclock'' experiments by Eckle {\i et al} (2008a,
2008b) \nocite{2008NatPhysEckle,2008ScienceEckle}.  In these
experiments,  timing of the tunneling ionization was mapped onto the
photoelectron momentum by application of an intense elliptically
polarized laser pulse. Such a pulse served both to liberate an
initially bound atomic electron and to deflect it in the angular
spatial direction. This deflection was taken as a measure of the
tunneling time.

\begin{figure}[ht]
\hs{1cm}
\bp{8cm}
\epsfxsize=15cm
\epsffile{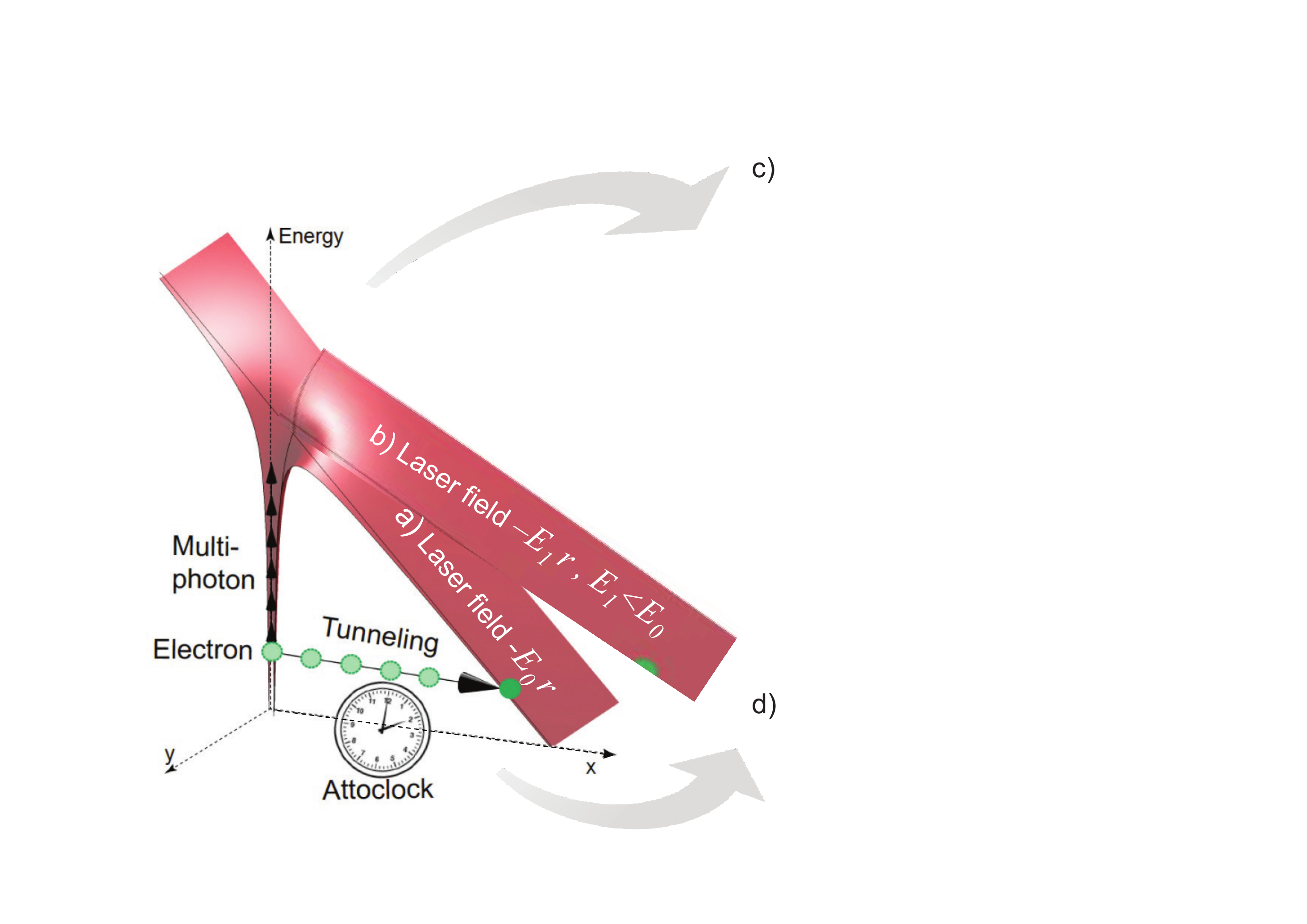}
\ep
\hs{0.9cm}
\bp{6cm}
\epsfxsize=6cm
\epsffile{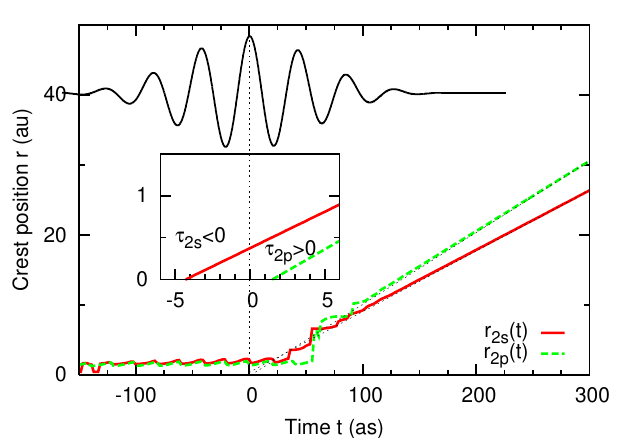}
\epsfxsize=6.8cm
\epsffile{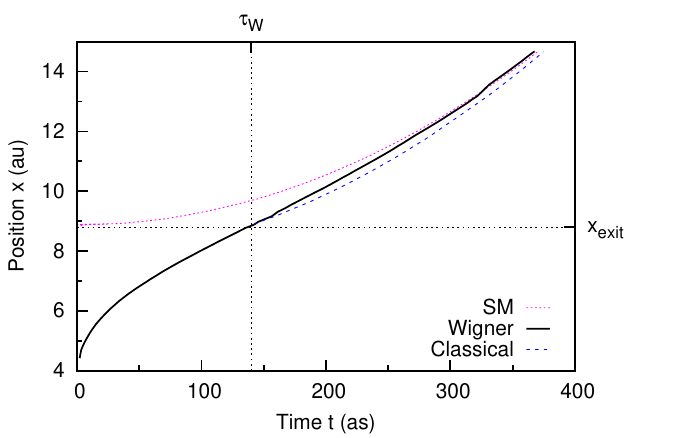}
\ep
\caption{ A man-made tunnel: a graphical illustration of the attoclock
  experiment (a) The laser field $E_0$ bends the atomic Coulomb
  potential and creates a penetrable potential barrier through which
  an initially bound electron tunnels out. Attoclock measures the time
  the electron spends under the barrier in a classically inaccessible
  region. Adapted from \cn{2015PhysRepLandsman}. (b) As the laser
  field decreases to $E_1<E_0$, the width of the barrier increases and
  this may increase the tunnelling time.  c) A Wigner time
  visualization by the photoelectron trajectory back-propagation in the
  multiphoton ionization regime.  Adapted from
  \cn{PhysRevLett.105.233002}.  d) The same trajectory visualization
  in the tunnelling ionization regime.  Data are from
  \cn{2017PRLCamus}.
\label{Fig1}
}
\end{figure}

The concept of tunneling time has received its first attention 
soon after the birth of quantum mechanics.  \cn{1932PhysRevMacColl}
considered a sub-barrier transmission of a one-dimensional wave packet
and concluded that
\begin{quote}
``\ldots there is no appreciable delay in the transmission of the packet
through the barrier.''
\enlargethispage{1cm}
\footnote{ \cn{1962JApplPhysHartman} showed that with the aid of
  greater computational power than available to
  \cn{1932PhysRevMacColl} that the same analysis and tunnelling time
  definition instead leads in general to non-zero values.  }
\end{quote}
This work triggered an intensive debate and the concept of tunnelling
time had been re-examined again under different guises. In a
summarizing review, \cn{RevModPhys.66.217} made the following remark:
\begin{quote}
``Over sixty years ago, it was suggested that there is a
  time associated with the passage of a particle under a tunnelling
  barrier. The existence of such a time is now well accepted; in fact
  the time has been measured experimentally. There is no clear
  consensus, however, about the existence of a simple expression for
  this time, and the exact nature of that expression \ldots''
\end{quote}
It is for this elusive nature of the tunnelling
time  that the first attoclock experiments by Eckle {\i et al} (2008a,
2008b) were so enthusiastically welcomed. The conclusion of these experiments
was resolute. Eckle {\i et al}  (2008a) unequivocally stated that
\begin{quote}
``Thus, the numerical simulations, like the
experiment, lead to a momentum distribution that
is consistent with a zero delay time for tunnelling.''
\end{quote}
A similar statement was made in the follow-up investigation by  the
same group \cite{2012NatPhysPfeiffer}:
\begin{quote}
``The excellent agreement of our theory for both atoms (Ar and He) and
  over a large intensity range below and above the Keldysh parameter
  $\gamma=1$ confirms zero tunnelling time within the experimental
  accuracy of 10~as.''
\end{quote}
These early experiments were conducted with relatively high laser
field intensities. In more recent experiments
\cite{PhysRevLett.111.103003,2014OpticaLandsman}, low field
intensities were accessed. As is seen from a schematic representation
of an attoclock measurement (\Fref{Fig1}~a-b), a weaker laser field
bends the Coulomb barrier less and hence the tunnel becomes wider in
a low field regime.  This possible increase of the tunnel width may
have resulted in a finite tunnelling time determination in the later
refined measurements
\cite{2014OpticaLandsman,2015PhysRepLandsman}. Following a very
thorough review and analysis of these experiments,
\cn{2019JModOptHofmann} concluded that
\begin{quote}
``\ldots  models including finite tunnelling time are consistent with recent
 experimental measurements''
\end{quote}
Thus, the whole decade of the attoclock experiments has ended
inconclusively keeping the door open for a finite tunnelling time.

Fresh fuel to the tunnelling time debate was added by a recent
measurement of \cn{2017PRLCamus} who titled their work
\begin{quote}
``Experimental evidence for quantum tunnelling time''.
\end{quote}
Not only did they detect a finite tunnelling time, but the values
attached to this time were very large, more than 100~as. These
values were derived by the Wigner trajectory  analysis. Such an
analysis is usually conducted to visualize the Wigner time that is
characteristic of strong field ionization in the multi-photon regime
\cite{M.Schultze06252010,PhysRevLett.105.233002}.
The Wigner time \cite{PhysRev.98.145} characterizes the photoelectron
group delay, or advancement, relative to the free space
propagation. It can be visualized by the back-propagation of the
photoelectron trajectory to the origin and its termination at a ``time
zero'' that is displaced relative to the peak electric field of the
driving laser pulse.  This displacement is the measure of the Wigner
time delay. \Fref{Fig1}c) illustrates the Wigner time determination by
back-propagating the photoelectron trajectories emitted from the $2s$
and $2p$ shells of the Ne atom. The inset of this figure clearly shows
the origin of these trajectories to be displaced to the opposite
directions relative to the peak of the driving pulse. Thus determined
Wigner time difference of the $2s$ and $2p$ shells of Ne is of the
order of 10~as. While the initial measurement by
\cn{M.Schultze06252010} valued this difference at $21\pm5$~as, a more
recent experiment by \cn{Isingerer7043} found it in a much closer
agreement with the theoretical predictions. As is seen from
\Fref{Fig1}(a-b), in the multi-photon ionization regime, the
photoelectron has enough energy to make a vertical transition and
emerges in the continuum close to the origin. In the tunnelling
ionization regime, the photoelectron makes a horizontal transition and
the tunnel exit point extends to many Bohr radii away from the origin
(\Fref{Fig1}~d). Termination of the photoelectron trajectory at such a
large distance results in a seemingly large ``Wigner'' tunnelling time
running over 100~as. However, such a Wigner-like definition of the
tunnelling time is questionable as a classical photoelectron
trajectory cannot be continued into a classically inaccessible region
under the barrier. Moreover, tunneling acts as an energy filter,
favouring higher-energy components of the photoelectron
wave packet and distorting the group delay \cn{2019JModOptHofmann}.

Meanwhile, contrary to some experimental evidence supporting a
finite tunneling time, a growing body of theoretical work points to a
vanishingly small or even zero tunnelling
time. \cn{2019JModOptHofmann} based their claim of a finite tunnelling
time on a sole helium measurement
\cite{PhysRevLett.111.103003,2014OpticaLandsman} that deviated
strongly from semi-classical modeling assuming instantaneous
tunnelling. However, this measurement did also deviate very
significantly from fully quantum simulations based on a numerical
solution of the time-dependent Schr\"odinger equation (TDSE)
\cite{PhysRevA.89.021402,Scrinzi2014}.  These \textit{ab initio}
simulations did not require any assumptions regarding tunnelling time
or adiabaticity scenario.  \cn{Scrinzi2014} termed this disagreement
as ``a very disquieting''. As a possible source of this disagreement,
\cn{Rost2019} pointed to an inconsistent field intensity calibration
which could affect a self-referencing attoclock measurement.

Another group of theoretical investigations advocating a zero
tunnelling time is not directly related to the performed experiment.
These numerical simulations, which have no laboratory counterparts and
that are termed for this reason {\i a numerical attoclock}, consider
atomic ionization driven by an ultra-short nearly single-optical-cycle
laser pulse. The photoelectron momentum distribution (PMD) in such a
field configuration is particularly simple. It can be simulated by
various simplified, but more physically transparent, techniques such
as an analytic $R$-matrix theory \cite{2015NatPhysTorlina}, a
classical back-propagation analysis \nocite{PhysRevLett.117.023002},
classical-trajectory Monte Carlo simulations
\cite{0953-4075-50-5-055602}, a classical Rutherford scattering model
\cite{2018PRLBray} and the strong field approximation (SFA)
implemented within the saddle point method (SPM)
\cite{PhysRevA.99.063428}. By making comparisons with these models, the
numerical attoclock firmly points to a vanishing tunnelling time
\cite{2015NatPhysTorlina,PhysRevLett.117.023002,2018PRLBray}.  It is
shown in these works that a noticeable deviation of the PMD from the
simple canonical momentum conservation picture is related to the
Coulomb field of the ion remainder. The same conclusion was reached in
a joint experimental and theoretical investigation on the atomic
hydrogen by \cn{2019NatSainadh} who made an upper bound estimate on
the tunnelling time not exceeding 1.8~as.  Similarly, in the molecular
hydrogen, such an estimate is under 10~as \cite{Wei2019}.
In a numerical attoclock setup on negative ions, with no Coulomb
drag on the photoelectron, the tunnelling time is also vanishing
\cite{2019PRADouguet}.

This growing body of evidence motivates us to reconsider the question
of whether an attoclock measurement can be interpreted in terms of  a
finite tunnelling time.  This question is considered
in detail in the following sections of this review article. Our
concluding remark is that the window for a non-zero `tunnelling time'
in the context of attoclock measurements on  simple atomic or molecular
targets appears to have essentially closed.

\begin{figure}[h!]
   \centering
   \includegraphics[height=7cm]
                   {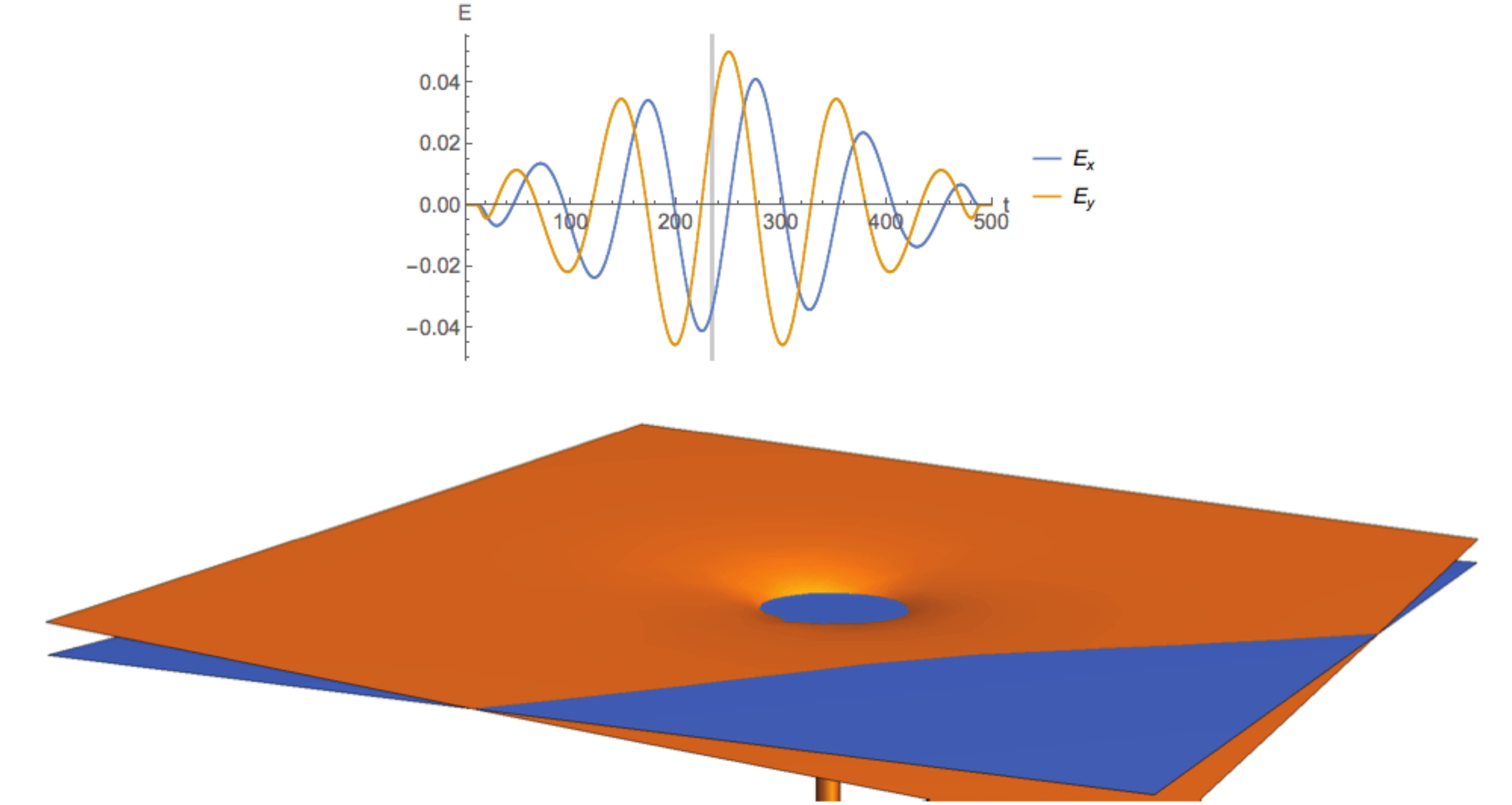}
   \caption{ The distortion of the atomic potential 
$ V_{\mathrm{tot}}(\bm{r},t)=1/r+\bm{E}(t)\cdot \bm{r}$
as a function of
     time for a typical attoclock pulse (orange surface) versus the
     initial binding energy (blue plane).  Where these two surfaces
     intersect the bound electron is able to adiabatically
     tunnel. Adapted from \cn{Bray2019}.  }
   \label{Fig2}
\end{figure}

\begin{figure}[ht]
\centering
\hspace{0.2cm} 
\includegraphics[height=5.27cm]
                {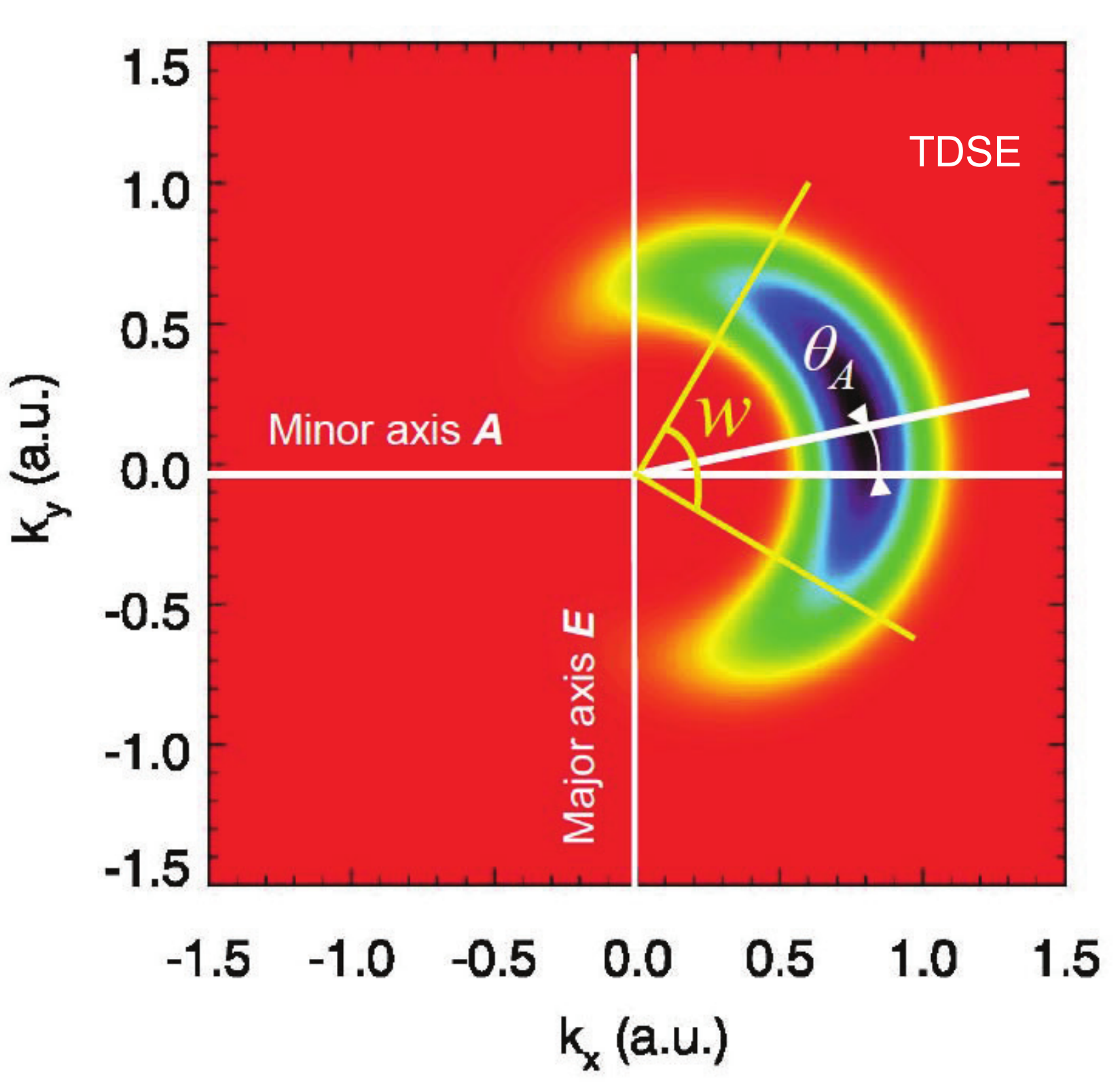}
\includegraphics[height=5.27cm]
                {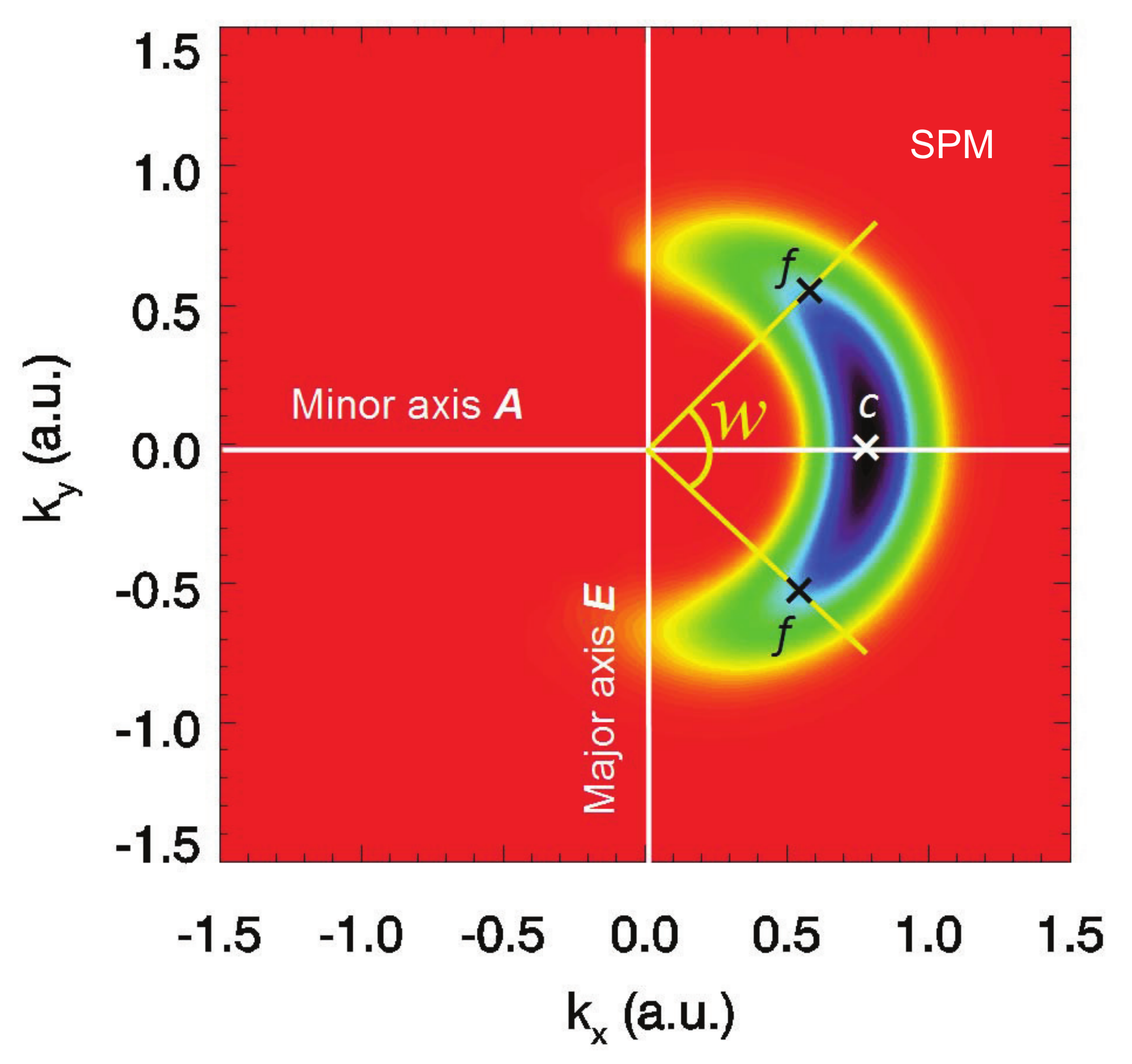}

\hspace{0.19cm}
\includegraphics[height=5.0cm]
                {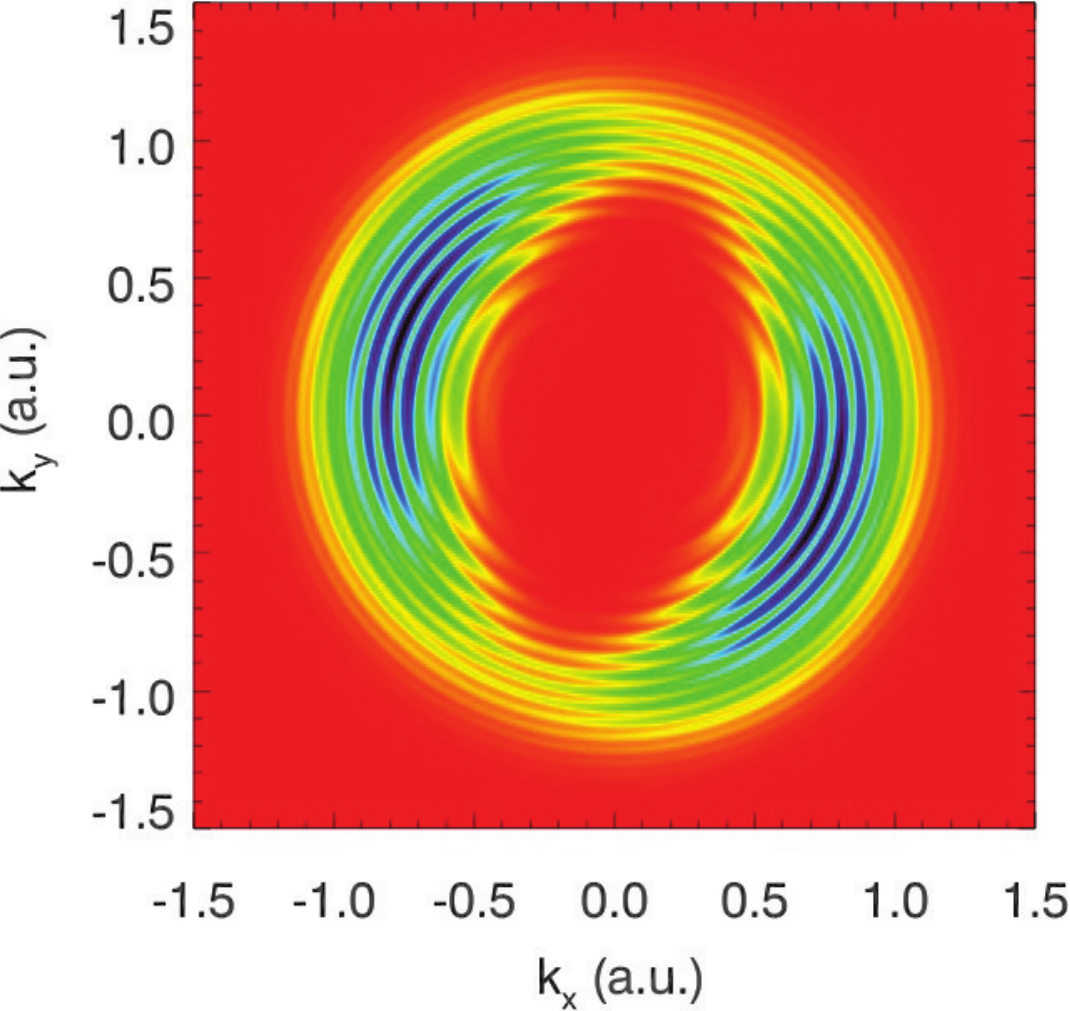}
\hspace{0.20cm}
\includegraphics[height=5.0cm]
                {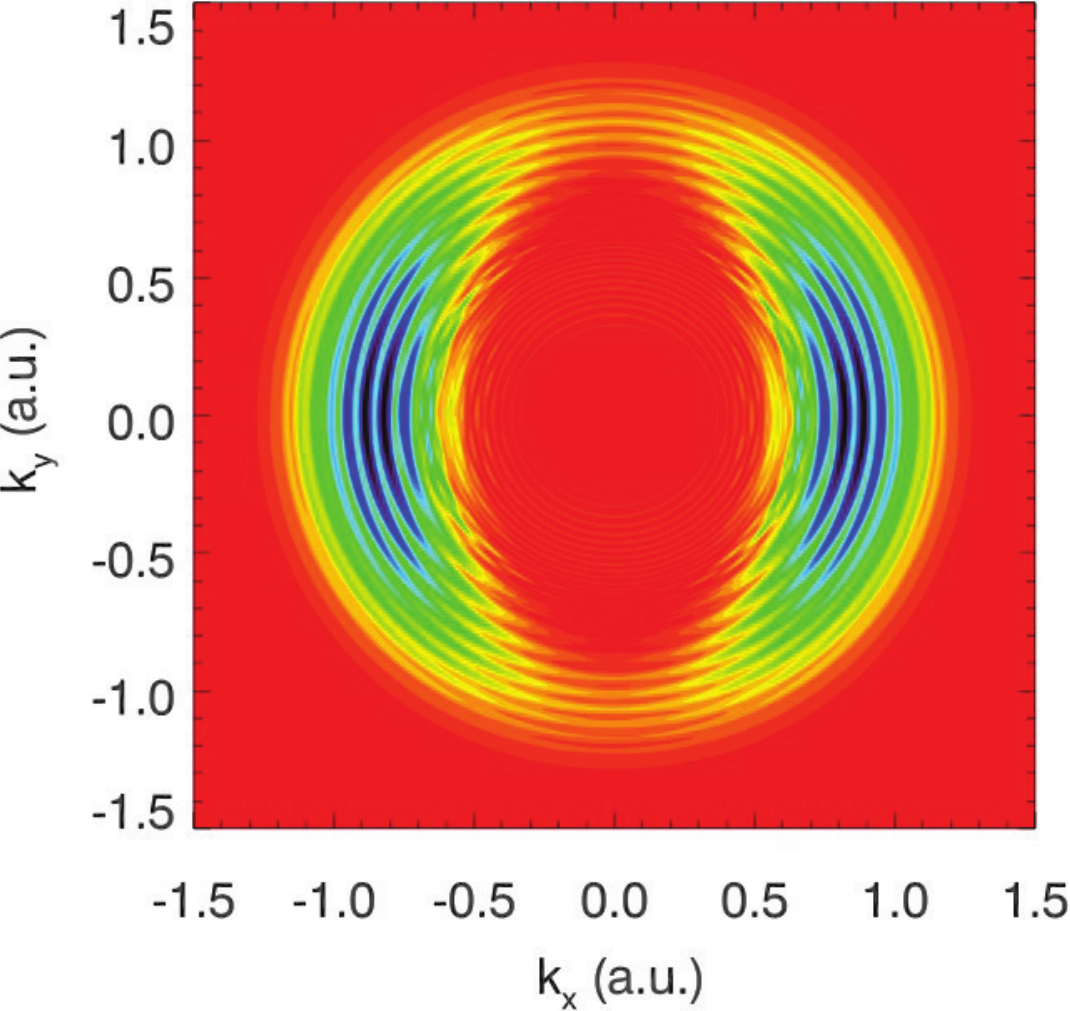}
\caption{Left column: TDSE calculated PMD of hydrogen for short (top)
  and long (bottom) pulses.  The short pulse is approximately 1.6 fs
  FWHM at the peak intensity of \Wcm{0.86}{14}, 800 nm, ellipticity
  $1.0$, and the anti-clockwise helicity.  The long pulse is
  approximately 6 fs FHWM at the peak intensity of \Wcm{1.5}{14}, 770
  nm, ellipticity $0.85$, and the clockwise helicity.  The PMD is
  averaged over CEP.  Right column: As for left but from a Yukawa
  potential (screening parameter $a=1$) with the hydrogenic binding
  energy.  The colouration of probability is linear and normalized
  from red (0.0) to black (1.0). Adapted from
  \cn{PhysRevA.99.063428} (top row) and \cn{Bray2019} (bottom row)}
\label{Fig3}
\end{figure}

\section{Attoclock principle}
\markboth{Attoclock principle}{Attoclock principle}

The attoclock employs the rotating electric-field vector $\bm{E}$ of
an elliptically polarized laser pulse to deflect tunnel-ionized
electrons in the angular spatial direction.  The field, that is
characterized by the temporal profile $f(t)$, the ellipticity
$\epsilon\lesssim 1$ and the magnitude $E_0$, is contained in the
polarization $(x,y)$ plane:
\be
E_x(t)= {E_0 f(t) \over \sqrt{1+\epsilon^2}}
 \cos(\omega t + \phi) \  , \
E_y(t)= {\epsilon E_0 f(t) \over \sqrt{1+\epsilon^2}}
 \sin(\omega t + \phi)
\ .
\label{ef}
\ee
 The instant of ionization $t_{\rm ion}$ is mapped onto the final
 momentum vector of the photoelectron at the detector
 $\bm{p}_{t\to\infty}$.  The tunneling ionization is an exponentially
 suppressed process. Predominantly, it commences at the time $t_0$
 corresponding to the peak of the driving laser pulse (see \Fref{Fig2}
 for illustration). At this instant, the vector-potential of the
 driving field
$
\bm{A}(t_0)=
-\int_{-\infty}^{t_0}{\bm E(t)}\ dt
$ 
is aligned with the minor $y$ axis of the polarization
ellipse \footnote[1]{We retain this nomenclature of the axes even for
  $\epsilon =1$ when the polarization ellipse becomes a circle.} .  It
is expected that the photoelectron emerges from the tunnel with zero
velocity (the adiabatic hypothesis) and its kinetic momentum captures
the vector potential of the laser field at the time of exit
$
\bm{p}_{t\to\infty}=-\bm{A}(t_{\rm ion})
\ .
$
Angular deviation of $\bm{p}_{t\to\infty}$ from the $y$ direction can be
simply related to the tunneling time $\theta = \omega\tau$, where
$\tau = t_{\rm ion} - t_0 \ $. The Coulomb field of the ion remainder
makes this interpretation less straightforward. Nevertheless, the
tunneling time can still be determined using a semi-classical trajectory
simulation. 

Any direct comparison of a fundamentally quantum mechanical process
with its classical analogue has many caveats. The attoclock principle
illustrates these caveats most vividly. 
Firstly, the PMD needs to be
characteristic of only one or few selected classical trajectories.  As
was shown by \cn{0953-4075-39-14-R01}, for close-to-circular
polarization, this number is exceeding by one the number of the
driving pulse oscillations. For a pulse with the temporal profile
\be
f(t)= \left\{
         \begin{array}{ll}
            \cos^4 (\omega t/2N)&-N\pi/\omega<t<+N\pi/\omega\\
            0&\mathrm{elsewhere}\\
         \end{array}
      \right.
\ ,
\label{envelope}
\ee
which becomes nearly single-cycle with $N=2$, there are only 3
contributing trajectories of which the one is strongly dominant
whereas the other two can be safely neglected.  Correspondingly, the
fully quantum mechanical simulation and the classical trajectory
simulation return very similar PMDs which both have a well defined
angle about which they are fully symmetric. See the top row of panels
in \Fref{Fig3} for the quantum-mechanical TDSE simulation (left) and
the semi-classical SPM simulation (right).
In contrast, for multi-cycle pulses with $N\simeq5$, the photoelectron
momentum distribution becomes less pronounced. It looses its natural
symmetry and acquires above-threshold-ionization (ATI) ring structure
from inter-cycle interference, a phenomenon unexplainable by classical
physics. Examples of such distributions are shown in the bottom row of
\Fref{Fig3} for the realistic atomic potential (left) and a
short-range Yukawa potential (right).

A further consideration pertinent to the observed attoclock momentum
distributions in \Fref{Fig3} is the effect of the carrier-envelope
phase (CEP) $\phi$ entering \Eref{ef}.  For short single-cycle pulses
(top row) the peak field strength simply rotates in the plane with
varying CEP and accordingly the same occurs for the resulting momentum
distribution.  However, for the few-cycle pulses with elliptical
polarization (bottom row), the direction of the peak field strength
only changes subtly with a variation of CEP.  In fact, it is only the
ellipticity that modulates the field strength significantly enough for
the distribution to exhibit the characteristic two-lobes observed in
the experiment, and for which none have been performed with stable CEP.
Consequently, these `long pulse' distributions are dependent on the
precise CEP and need to be accordingly averaged over for comparison
with a  CEP variable experiment.

\section{Attoclock  interpretation}
\markboth{Attoclock  interpretation}{Attoclock  interpretation}

\begin{figure}[h!]
\includegraphics[height=6.5cm]
                {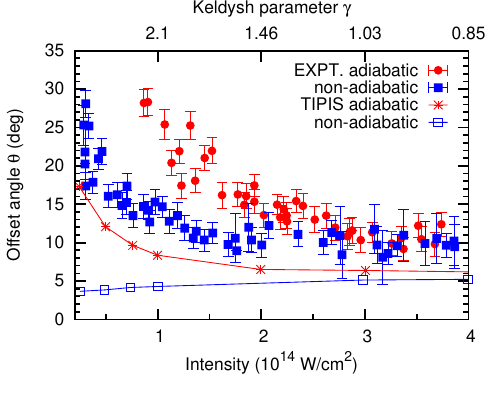}
\includegraphics[height=6.5cm]
                {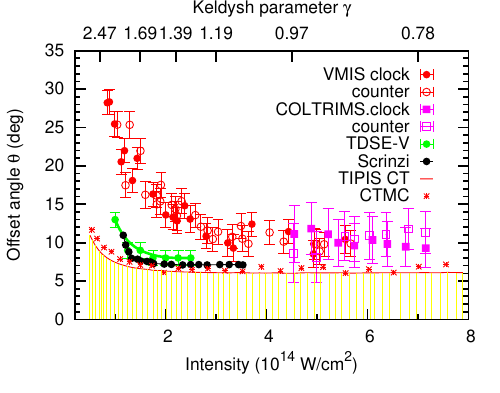}
\caption{Left: Experimental data and TIPIS simulations from
  \cn{PhysRevLett.111.103003} are shown along with the TDSE
  calculations by \cn{PhysRevA.89.021402} and  \cn{Scrinzi2014}.
Right: Experimental data and TIPIS simulations from
  \cn{2014OpticaLandsman} are compared with the same TDSE
  calculations. }
\label{Fig4}
\end{figure}

The second caveat of relating an attoclock measurement to a classical
trajectory is the appropriate boundary condition. This typically
depends on the assumption of the adiabaticity of the tunneling
process, i.e. whether the photoelectron emerges from the tunnel with
the zero (adiabatic) or a non-zero (non-adiabatic) velocity. This
adiabaticity scenario, in turn, affects the intensity calibrations that
have been in use \cite{hofmann2016non}.
In the primary experimental work (Eckle {\i et al} 2008a, Pfeifer {\i
  et al} 2012) \nocite{2012NatPhysPfeiffer} the comparison was
performed against a model known as tunnel ionization in parabolic
coordinates with induced dipole and Stark shift (TIPIS). 
This comparison led to the conclusion of zero tunneling time as there
was no discernible difference between experimental measurements and
the TIPIS predictions. The latter assumed  instantaneous tunneling.
In the later work, however, \cn{PhysRevLett.111.103003} and
\cn{2014OpticaLandsman} found quite the opposite, with the observed
offset angle being much larger than that predicted by TIPIS, and
attributed this difference to a finite tunneling time.  This
comparison is illustrated in \Fref{Fig4}. In the left panel of the
figure, the two experimental data sets of \cn{PhysRevLett.111.103003}
are shown which are calibrated on the intensity scale under the
adiabatic (red) and non-adiabatic (blue) tunneling scenarios. The two
analogous TIPIS calculations are also displayed. While the adiabatic TIPIS
simulation (red) is in qualitative agreement with the experiment, the
non-adiabatic TIPIS (blue) should be discarded as clearly unphysical.
A noticeable  difference between the adiabatic TIPIS and the adiabatic
experiment is wholly attributed to the tunneling time.

In the same panel, the two TDSE calculations are shown in black
\cite{Scrinzi2014} and green \cite{PhysRevA.89.021402}. These TDCS
calculations are much closer to the blue {\i non-adiabatic} set of
experimental data even though the non-adiabatic scenario was discarded
in the experiment because of the TIPIS simulation failure. It should
be also emphasized that the TDSE simulations were fully {\i ab initio}
and no  specific tunneling scenario was adopted. The only
approximation used in both TDSE calculations was the single active
electron of the helium atom interacting with the laser field. However,
the later and more refined simulation including both active electrons found
no electron correlation effects in the He attoclock setting
\cite{2017JModOptMajety}.

In the right panel of \Fref{Fig4}, an extended data set of
\cn{2014OpticaLandsman} is displayed. It contains only the adiabatic
data which are shown in red. The experiment is conducted both with the
cold target recoil-ion momentum spectroscopy (COLTRIMS, displayed with
squares) and a velocity map imaging spectrometer (VMIS, displayed with
circles). The experimental data are separated into the clockwise
(filled symbols) and the anti-clockwise (open symbols) helicity.
The adiabatic TIPIS simulations are carried over with the
single trajectory (CT, solid line) and classical trajectory Monte
Carlo (CTMC, asterisks) simulation. The shaded area below the CTMC
curve is the offset angle admissible under the zero tunneling time
scenario. The difference between the experiment and the TIPIS model,
which is barely noticeable at large field intensities (COLTRIMS), increases
significantly in the low-intensity regime (VMIS). This difference, in
principle, could  be attributed to a finite tunneling time. However,
more likely, it is due to inconsistent intensity calibration. Indeed, the
experiment deviates similarly strongly from the {\i ab initio} TDSE
calculations which do not assume any tunneling hypothesis or
adiabaticity scenario.

The latest attoclock measurement that suggested a non-zero tunnelling
time was reported by \cn{2017PRLCamus}. They adopted  theoretical
modeling of \cn{PhysRevA.90.012116} and evaluated a Wigner trajectory
using a fixed-energy propagator calculated from the phase of the
solution of the {\i time-independent} Schr\"odinger equation. Such a
quasi-stationary approach could only be applied in the strongly
adiabatic regime when the energy of the tunneling particle would be
conserved. After the quantum Wigner trajectory was found, it was matched
at the exit from the tunnel with a classical trajectory with the
initial conditions given by the Wigner formalism $p_{\rm exit} = p_W$
and $t_{\rm exit}= \tau_W$. The latter quantity was adopted as a
tunneling time. In addition to a classical trajectory determined by
the Wigner initial conditions, \cn{2017PRLCamus} defined a so-called
simple-man (SM) trajectory in which the photoelectron exits the tunnel
instantaneously $t_{\rm exit}=0$ and fully adiabatically $p_{\rm
  exit}=0$.
The Wigner, classical and SM trajectories are exhibited in
\Fref{Fig1}d) for the Kr atom at the field intensity $I=$\Wcm{1.7}{14}.

\begin{figure}[h!]
\hs{-0.7cm}
\includegraphics[height=5.1cm]
                {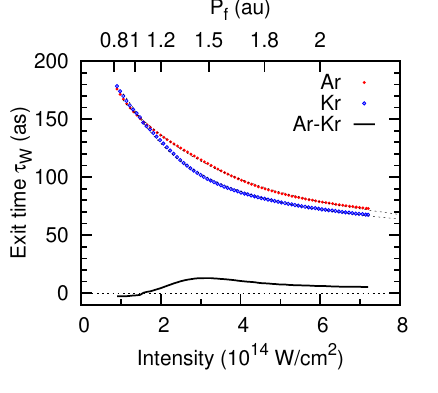}
\hs{-0.6cm}
\includegraphics[height=5.1cm]
                {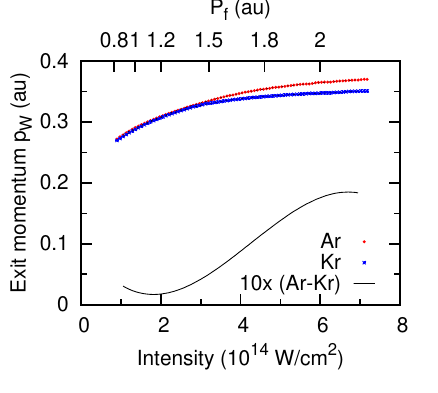}
\hs{-0.6cm}
\includegraphics[height=5.1cm]
                {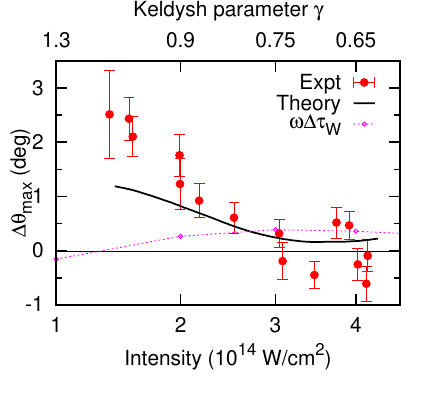}
\vs{-10mm}

\caption{Left and center: the exit time $\tau_W$ and the exit momentum
  $p_W$ in Ar (red dots) and Kr (blue dots). Their differences
  $\Delta\tau_W$ and $\times10\Delta p_W$ are shown solid black
  lines. Right: the experimental observable $\Delta\theta$ and its
  theoretical estimate, both the original data from \cn{2017PRLCamus},
  are plotted versus the laser field intensity. The Wigner component
  $\delta\theta_W=\w\Delta\tau_W$ is also shown.}
\label{Fig5}
\end{figure}

We analyze the results of \cn{2017PRLCamus} and the follow up work
by \cn{Camus2018} in more detail in \Fref{Fig5}.
Neglecting the ionic Coulomb potential in the SM model leads to the
final momentum aligned perfectly with the minor polarization axis and
having the zero offset angle $\delta\theta=0$. At the same time, for the
Wigner trajectory, the additional time delay $\tau_W$ manifests itself
as a rotation of the asymptotic momentum distribution by
$
\delta\theta_{\tau} = \w\tau_W
$
and the nonzero initial momentum $p_W$ to a counter-rotation by
$
\delta\theta_p \approx -p_W/p_E
$
with the drift momentum being
$
p_E=E_0/\w
\ .
$
Here $E_0=\sqrt{I/[I_0(1+\epsilon^2)]}$ while $\epsilon$ is the
ellipticity and $I_0=$~\Wcm{3.51}{16} is the atomic unit of the field
intensity.

The exit times $\tau_W$ and the exit momenta $p_W$ for Ar and Kr as
functions of the field intensity are exhibited in the left and central
panels of \Fref{Fig5}, respectively. Their differences are calculated
from the original data of \cn{2017PRLCamus} and  also plotted. The
right panel displays the final set of the experimental data in
which the difference in the attoclock offset angles $\Delta\theta_{\rm
  max}=\theta^{\rm Ar}_{\rm max}- \theta^{\rm Kr}_{\rm max}$ is
compared with the predictions of the theory. Both the theory and
experiment are presented by \cn{2017PRLCamus} on the momentum scale
$
\rho^{\rm Ar}(\theta_{\max}) =
\sqrt{
p_x^2(\theta_{\max})+
p_y^2(\theta_{\max})
}
\ .
$
We assume that $\rho(\theta_{\max})\approx \sqrt{p^2_W+p^2_E}$ and
place the $\Delta\theta_{\rm max}$ data on the absolute intensity
scale using the $p_W$ values shown in the central panel of
\Fref{Fig5}. This allows us to plot on the same graph the Wigner time
induced component $\Delta\theta^W_{\max}=\w\Delta\tau_W$ calculated
from the Wigner times difference shown in the left panel. The results
are very indicative.  The essence of every clock is to tell the
time. We observe, however, that the time induced rotation difference
is vanishingly small in this experiment. The same can be said about
the initial momentum induced component. Indeed, the difference in
$p_W$ between Ar and Kr is only noticeable at larger intensities. At
lower intensities, where $\Delta\theta_{\max}$ is largest, it is
induced neither by $\Delta\tau_W$ nor $\Delta p_W$ but solely by the
different effect of the photoelectron scattering in the Coulomb field
of the ion reminder. Indeed, as \cn{2017PRLCamus} pointed out, the
attoclock set up is very sensitive to the tunnel width. This width is
estimated in the Keldysh theory as $x_{\rm exit}=I_p/E_0$
\cite{Keldysh1965}. The difference in the ionization potentials
between Ar and Kr causes the difference in $x_{\rm exit}$ which
becomes greater in the lower field intensity as $E_0$ decreases. This is
accompanied by the increase in $\Delta\theta^W_{\max}$. Thus, we may
say that the attoclock is, in fact, a ``nano-ruler'' that provides a
very accurate determination of $x_{\rm exit}$. It becomes ever more
sensitive in the weaker field regime.

\section{Hydrogen versus noble gas atoms}
\markboth{Hydrogen vs noble gases}{Hydrogen vs noble gases}

Some uncertainty in interpretation of the attoclock measurements on
helium \cite{PhysRevLett.111.103003,2014OpticaLandsman} and heavier
noble gas atoms \cite{2017PRLCamus} could possibly be attributed to
the effect of many-electron correlation. Indeed, the initial
theoretical modeling \cite{PhysRevA.89.021402,Scrinzi2014}, which was
found at variance with the helium experiment, was conducted in the
single active electron approximation while the effect of correlation
was neglected. Such a correlation could, in principle, slow down the
tunneling process. The tunneling electron may need the time to
negotiate with its many-electron environment the fine detail of the
tunneling process. However, in the case of He, this was explicitly
shown not to be the case \cite{2017JModOptMajety}. Indeed, the
remaining electron in the He$^+$ ion is so tightly bound that it
hardly interacts with the tunneling electron other than by screening the
charge of the bare nucleus. This situation may differ for
such complex atoms as Ar and Kr where a precise theoretical modeling
with an accurate account for many-electron correlation is not feasible
at present.

The hydrogen atom is naturally correlation free. Hence, an attoclock
measurement on this  atom would give the cleanest determination
of the tunneling time.  However, until very recently, such a
measurement was not possible because of a very low density of atomic
hydrogen targets and hence a low count rate in the COLTRIMS
setting. Finally, this experimental hurdle was overcome and the first
hydrogen attoclock experiment was reported by \cn{2019NatSainadh}.

\begin{figure}[h!]
\hs{-0.7cm}
\includegraphics[height=6.1cm]
                {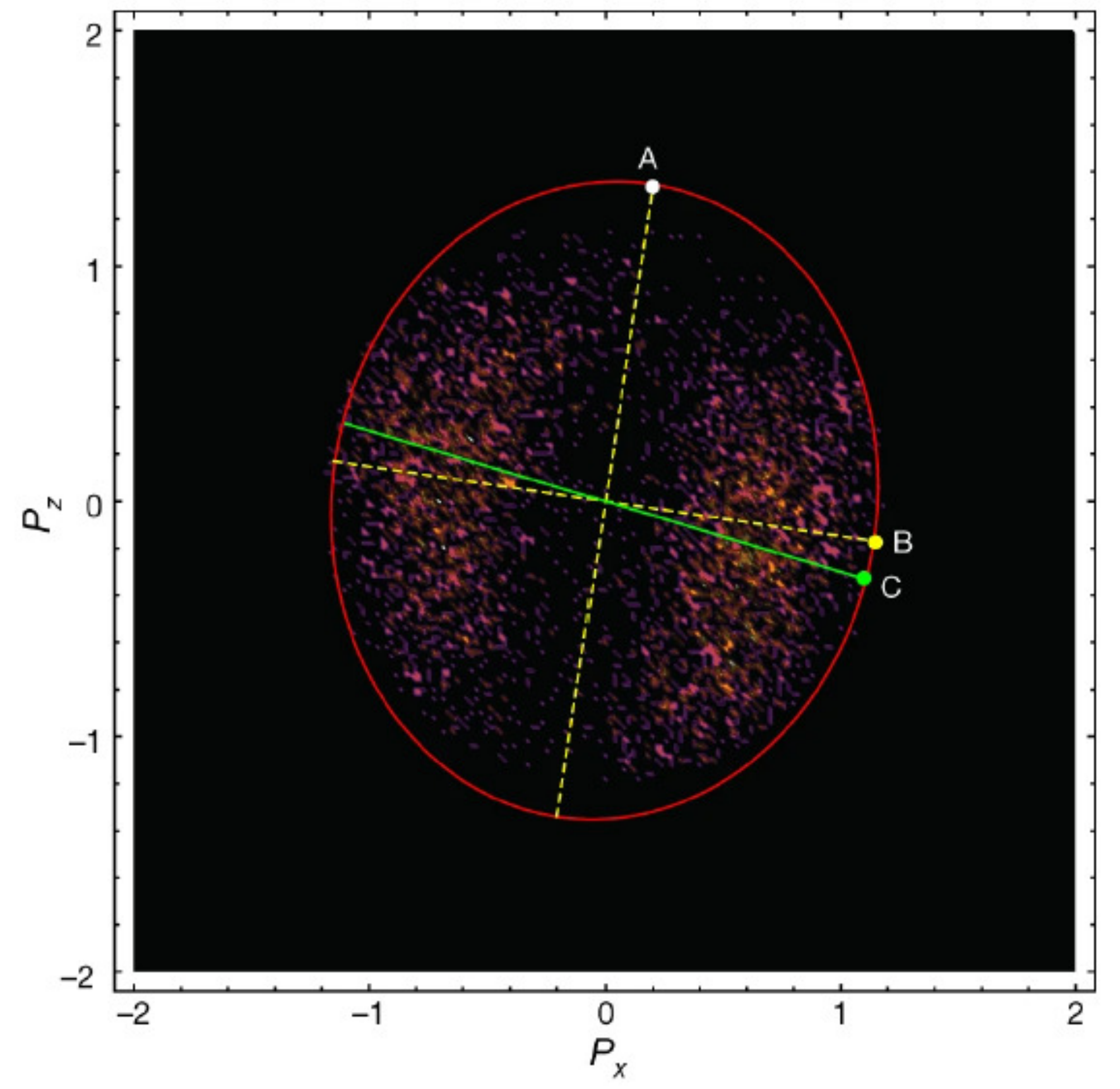}
\hs{0.6cm}
\includegraphics[height=6.1cm]
                {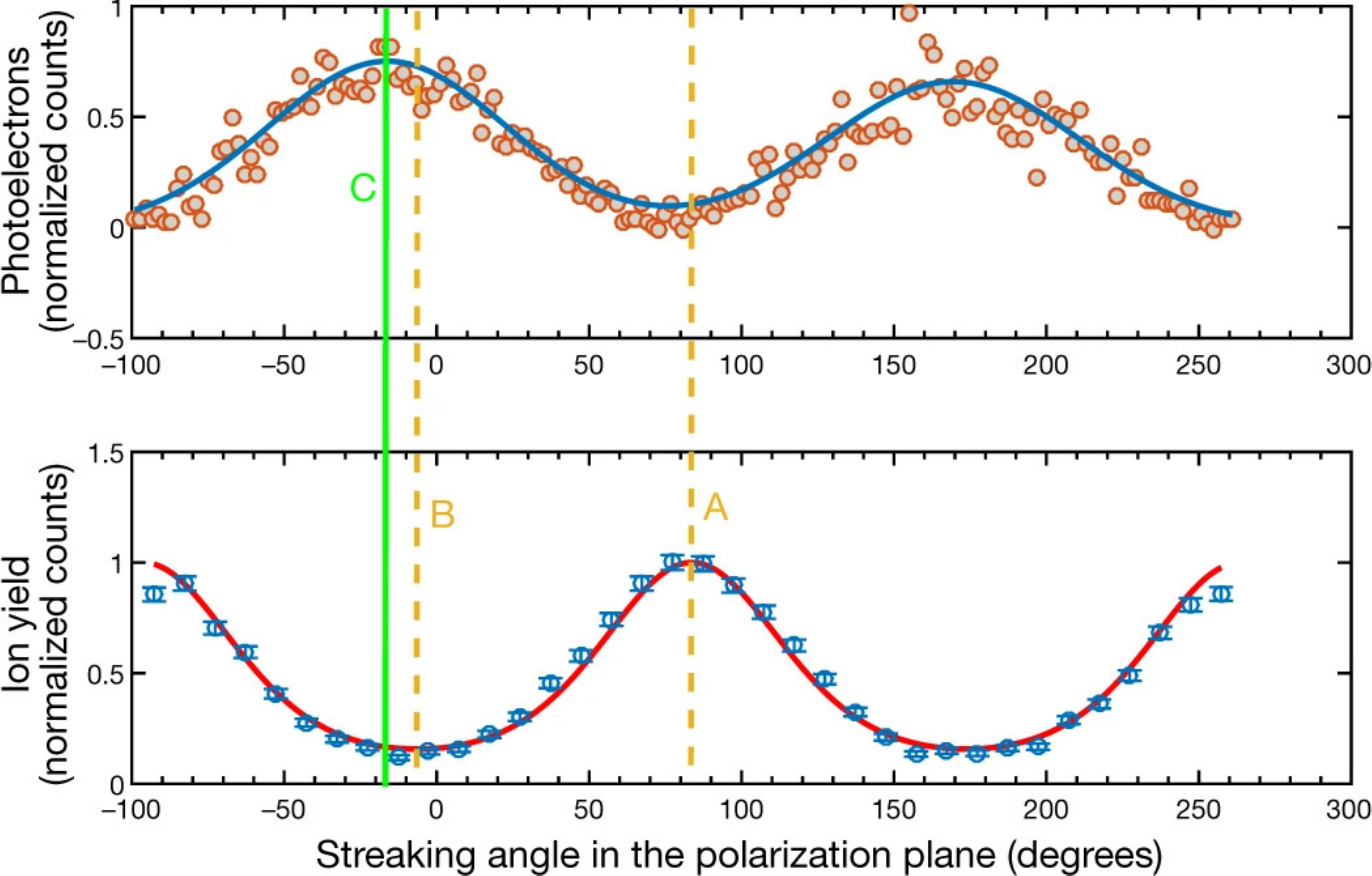}

\caption{Left: PMD of the hydrogen atom is projected on the
  polarization plane. Points A and B mark the direction of the major
  and minor axes of the polarization ellipse. Right: radially
  integrated counts of the photoelectrons (top) and the photo-ions (bottom). Point
  C marks the angular maximum of the photoelectrons count. Adapted
  from \cn{2019NatSainadh}}
\label{Fig6}
\end{figure}

Experimental results of \cn{2019NatSainadh} are illustrated in
\Fref{Fig6}. The left panel displays the photoelectron momentum
distribution projected onto the polarization plane. The major A and
minor B axes of the polarization ellipse are aligned with the electric
field $\bm E_0$ and the vector potential $\bm A_0$ at the instant of
the tunneling. In the SM picture, under the conditions of $\tau_{\rm
  exit}=0, p_{\rm exit}=0$ and neglecting the Coulomb field of the ion
remainder, the PMD should peak in the B direction. However, this peak
is displaced. This displacement, taken as the attoclock offset angle
$\theta_A$, is clearly seen in the right panel of the figure. Here the
radially integrated counts of the photoelectrons (top) and the
photo-ions (bottom) are shown. The ion counts are obtained with the
linearly polarized light, a precursor of the elliptical light, to mark
most accurately the major polarization axis direction.

\begin{figure}[h!]

\centering
\includegraphics[height=9cm]
                {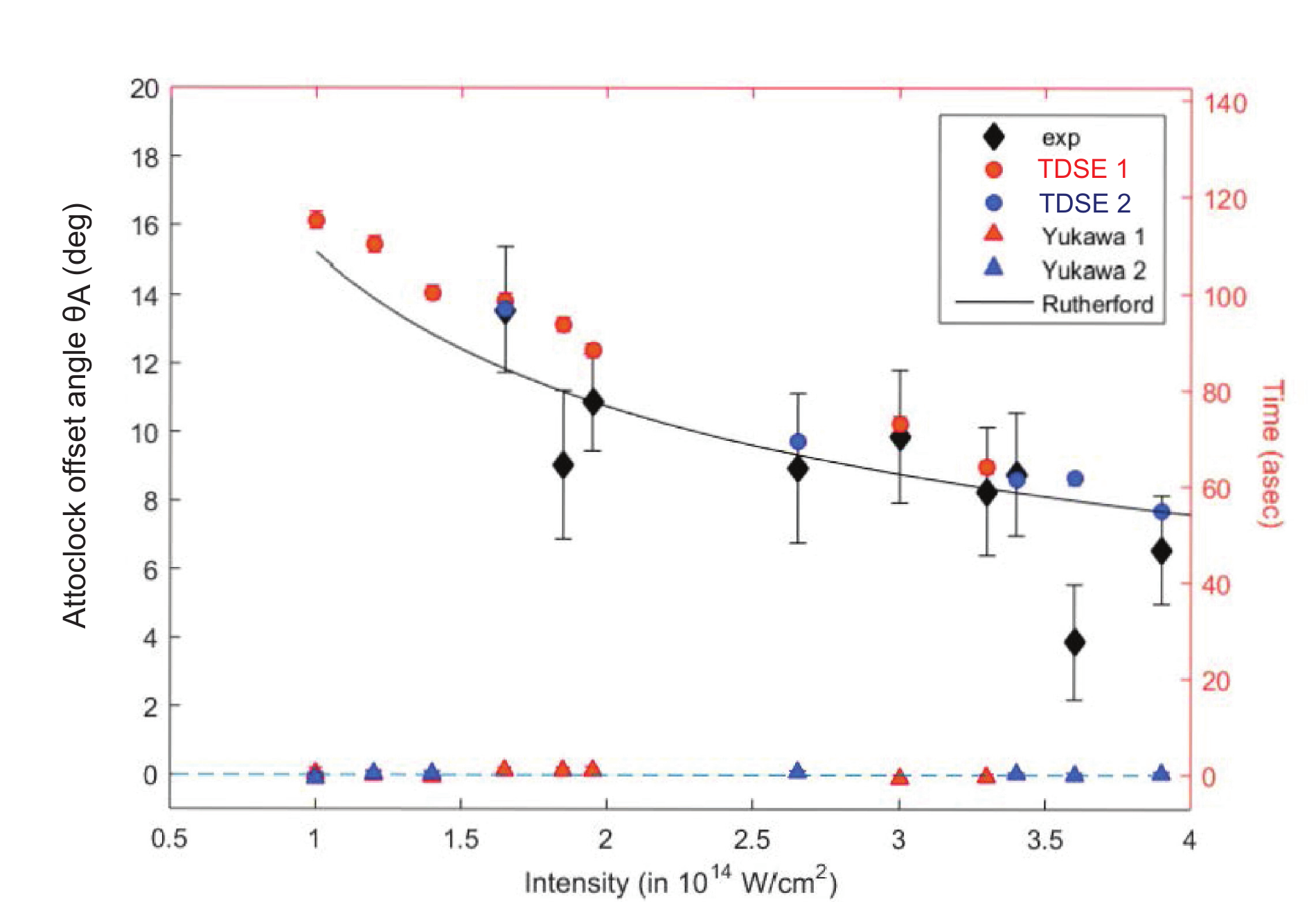}

\caption{The attoclock offset angle $\theta_A$ as a function of the
  laser pulse intensity is extracted from the experiment and
  calculated by two TDSE1 and TDSE2 codes. Analogous calculations with
  a model Yukawa atom are marked as Yukawa 1 and 2.  The intensity
  scaling $I^{-1/2}$ as prescribed by the classical Rutherford
  scattering model \cite{2018PRLBray} is also marked.  Adapted from
  \cn{2019NatSainadh}}
\label{Fig7}
\end{figure}

The attoclock offset angles $\theta_A$ as a function of the laser
field intensity are plotted in \Fref{Fig7}. The experimental values
are extracted from the radially integrated photoelectron momentum
density projected onto the polarization plane (see \Fref{Fig6} for
illustration). In the same \Fref{Fig7}, results of the two
calcualtions are also shown. These calculations, marked as TDSE1 and
TDSE2, utilized two computer codes developed independently by
\cn{PhysRevA.93.033402} and \cn{PhysRevA.90.013418}.  To make the
closest possible comparison with the experiment, the calculated
PMDs were processed in exactly the same way as the
experimental data. They were projected on the polarization plane and
radially integrated by the same numerical algorithm.  As the CEP was
not stabilized in the experiment, the calculated data were also
averaged over CEP values ranging from 0 to $2\pi$ in steps of
$\pi/4$. This procedure led to a rather satisfactory agreement between
theory and experiment which validated the experimental technique and
the two numerical TDSE solutions.

To eluscidate the long-range Coulomb field effect, a model Yukawa atom
was constructed with a screened Coulomb potential $U_Y (x) = -Z/r
e^{-r/a}$ with $a=1$ and $Z=1.908$. The binding energy in such a
potential is the same as in the hydrogen atom. The simulated PMDs in
the polarization plane are shown in the bottom row of panels in
\Fref{Fig3} (hydrogen - left, Yukawa - right). In these simulations,
the experimental laser pulse parameters are adopted with the FHWM of
approximately 6 fs, the peak intensity \Wcm{1.5}{14}, the wavelength
of 770 nm and the ellipticity $0.85$ as in \cn{2019NatSainadh}. The
offset of the peak PMD relative to the minor polarization axis,
clearly seen in the hydrogen atom, all but disappears for its
Yukawa counterpart. The TDSE 1 and 2 calculations with the Yukawa
potential, marked as Yukawa 1 and 2 in \Fref{Fig7}, produce
vanishingly small angular offsets at all laser pulse intensities. The
error bars in these Yukawa calculations, resulting from a Gaussian fit
to the radially integrated angular distributions, are not exceeding
$\delta\theta_{\rm Yukawa}=0.25^\circ$ which corresponds to a maximum
possible time delay of $\tau_{\rm Yukawa}=\delta\theta_{\rm
  Yukawa}/\w=1.8$~as.  \cn{2019NatSainadh} has taken this number as
the upper bound of the tunneling time in hydrogen. Indeed, as the
binding energy and hence the tunnel width are identical in the
hydrogen and Yukawa atoms, their tunneling times should be also close.

\section{Numerical attoclock}
\markboth{Numerical attoclock}{Numerical attoclock}

Prior to the experiment of \cn{2019NatSainadh}, several attempts have
been made to evaluate the tunneling time in the hydrogen atom by
conducting simulations with ultra-short, nearly single-oscillation
pulses.  As we outlined in the introduction, these ``numerical
attoclock experiments'', which could not be matched by laboratory
measurements, have a great utility of a very transparent physical
interpretation provided by several simplified analytical or numerical
tools calibrated against {\i ab initio} TDSE solutions.  In the
following sections, we give a brief review of these tools.

\subsection{Analytical $R$-matrix theory}

In an analytical $R$-matrix theory (ARM), the probability of detecting
an electron with a certain momentum is described by a time integral
over all possible instances of ionization \cite{PhysRevA.86.043408}.
This integral is expressed in terms of quantum trajectories whose
contribution is evaluated using the saddle-point method. The saddle
point equations return the starting times of the photoelectron
trajectories which turn out to be complex \cite{MIvanov2005}. Their
real part is taken as the ionization time and it is counted relative
to the peak of the driving laser pulse. The ionization time corresponding
to detection of a photoelectron with a momentum $p$ at an angle $\phi$
in the polarization plane is expressed as
\be 
\label{time}
t_i(p,\phi)
\equiv 
{\rm Re} t_s(p,\phi)
={\phi\over\w} 
+ \Delta t^{\rm env}_i(p,\phi)
+ \Delta t_i^C
\ \ , \ \ 
 \Delta t_i^C = 
- {dW_C(\phi,p)\over I_p}
\ee
Here $W_C$ is the phase acquired by the laser-driven electron due to
its interaction with the ionic core and $I_p$ is the ionization potential. A
small correction $\Delta t^{\rm env}_i$ is due to the ultrashort pulse
envelope.

\begin{figure}[h!]

\hs{-0.5cm}
\includegraphics[height=5.6cm]
                {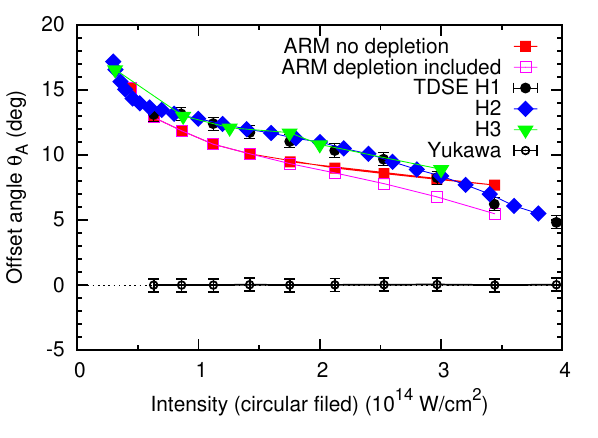}
\hs{-0.5cm}
\includegraphics[height=5.6cm]
                {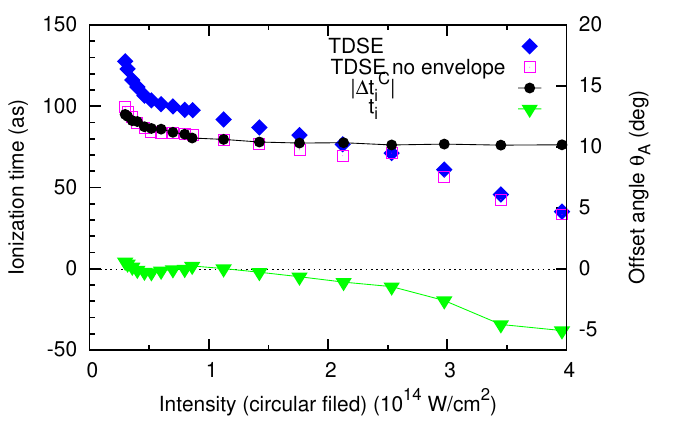}
\hs{-1.5cm}

\caption{Left: the attoclock offset angles $\theta_A$ from the
  analytical ARM (with and without account for depletion) and
  numerical TDSE (H1, H2, H3) calculations on the hydrogen atom. A
  TDSE calculation on the model Yukawa atom is also shown. Right: TDSE
  H2 calculation is converted to the tunneling time with and without
  envelope correction. The Coulomb correction $|\Delta t_i^C|$ is
  shown separately and subtracted from the envelope corrected TDSE to
  give the net tunneling time $t_i$. The data are from
  \cn{2015NatPhysTorlina}}
\label{Fig8}
\end{figure}

\cn{2015NatPhysTorlina} calibrated the ARM against numerically exact
TDSE solutions. The TDSE solution for the hydrogen atom can be found
exactly within the standard nonrelativistic and dipole
approximations. The ARM calibration is illustrated in the left panel
of \Fref{Fig8} where the attoclock offset angles $\theta_A$ are
plotted for various laser pulse intensities. The figure displays two
sets of the ARM results. The depletion of the target atom is included
in one set and its effect becomes noticeable at larger field
intensities. The comparison is also made with three sets of TDSE
calculations utilizing independently developed computer codes
\cite{Muller1999,Tao2012,PhysRevA.89.021402}. Agreement between the
TDSE results is very close and it can be used to benchmark the ARM
calculations. After the numerical accuracy of the ARM is established,
it can be used for extracting the ionization times. For this purpose,
\Eref{time} is solved for the $p,\phi$ values corresponding to the
peak PMDs obtained from TDSE calculations (see the bottom left panel
of \Fref{Fig3} as an example).  The ionization times extracted from
the ARM and TDSE calculations are plotted in the right panel of
\Fref{Fig8}. Here the TDSE H2 offset angles are converted to the
ionization times using the first term in \Eref{time}. By adding the
second term, an envelope-free TDSE result is obtained.  When both the
envelope correction and the Coulomb correction $|\Delta t_i^C|$ (
shown separately in the figure) are subtracted, the ionization times
become very small. They deviate noticeably from zero at larger field
intensities when the depletion effect becomes strong. This deviation
to negative ionization times mean that those photoelectrons arrive to
the detector that started tunneling before the laser pulse reached its
peak value. When this peak value is reached, the target atom is
already depleted. This observation challenges one of the core
assumptions of the attoclock measurement.  Based on the exponential
sensitivity of strong-field ionization to the electric field, it is
assumed that the highest probability for the electron to tunnel is at
the peak of the electric field. It turns out that this assumption is
not always correct. Meanwhile, under no condition is the ionization
time positive, i.e. traversing the potential barrier does not take any
real time.

\subsection{Classical back-propagation}

Ni and co-authors \cite{PhysRevLett.117.023002,PhysRevA.97.013426}
proposed a novel computational scheme which allowed them to determine
the time and position of the electron exiting the tunnel in strong
field atomic ionization. The scheme involves a quantum propagation of
the ionized electron for a sufficiently long time and its return, by
back-propagation, to the point of exit along various classical
trajectories. Only those trajectories are accepted that pass near
the origin with close to zero velocities, i.e. if the kinetic
momentum in the instantaneous field direction vanishes.  Typically
this condition is fulfilled several times along a trajectory, and
the event closest to the ion is identified as the tunnel exit.
The ensemble of the tunneling ionization trajectories determines a
distribution of the tunneling times ${\mathcal P}(\tau)$ relative to
the peak position of the driving laser pulse. The mean tunneling time
and the total tunneling probability are calculated as 
$
\la \tau \ra = \int d\tau \tau d{\mathcal P}/d\tau
$
and
$
P_{\rm tun} = \int d\tau {\mathcal P}(\tau)
\ , 
$
respectively. Meanwhile, the quantum propagation of the photoelectron
wave function for a sufficiently long time allows one to find the total
ionization probability $P_{\rm ion}=\int|\Psi|^2 d\r$. By knowing both
$P_{\rm ion}$ and $P_{\rm tun}$, the fraction of
not-tunneling ionization events can be expressed as  
$
\chi = (P_{\rm ion}-P_{\rm tun})/P_{\rm ion}
\ .
$

\begin{figure}[h!]

\hs{-0.cm}
\includegraphics[height=5.6cm]
                {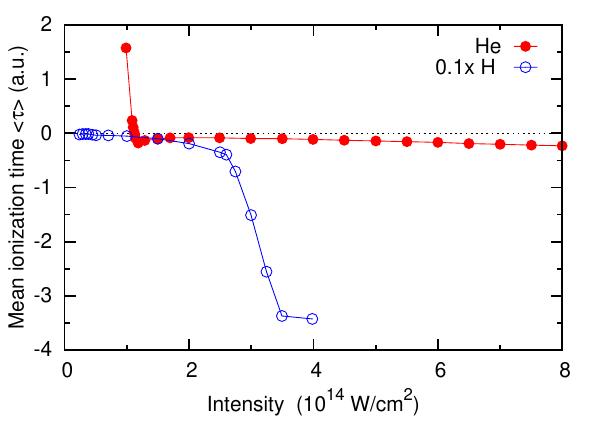}
\hs{-0cm}
\includegraphics[height=5.6cm]
                {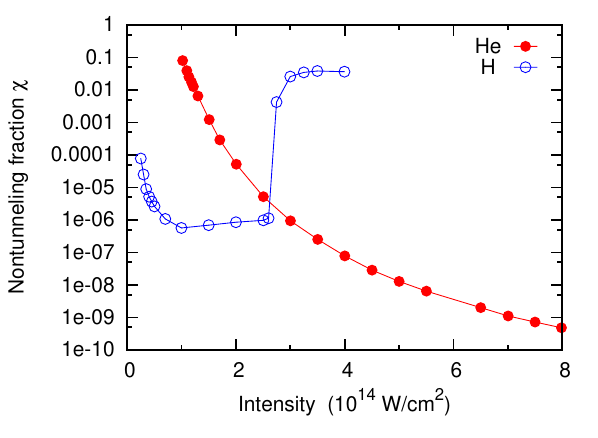}
\hs{-1.5cm}

\caption{Left: The mean ionization time $\la\tau\ra$ in He and H atoms
  as a function of the laser field intensity. Right: The fraction of
  non-tunneling ionization events $\chi$ in He and H across the same
  field intensity range.  The data are from
  \cn{PhysRevLett.117.023002}}
\label{Fig9}
\end{figure}

Results of the tunneling time determination for hydrogen and helium
atoms are shown in the left panel of \Fref{Fig9}. The mean tunneling
time $\la \tau\ra$ is close to zero for both target atoms when the
fraction of non-tunneling ionization events is small. This fraction
grows for both targets at the edges of the considered laser pulse
intensity range. In helium, because of its larger ionization potential,
this fraction becomes significant in the low-intensity range at the
onset of the multi-photon ionization regime. We note that $\gamma=2$
for He at $I=$\Wcm{1}{14}. The same increase in $\chi$ is seen at low
intensities for H as well, however, it is not that strong and not
exceeding $10^{-4}$. At the same time, because of a lower ionization
potential, the depletion effect is much stronger for H and it affects
the fraction of non-tunneling ionization events at the higher intensity
range. Here the mean tunneling time becomes strongly negative. The
same effect was observed by \cn{2015NatPhysTorlina} and can be seen in
\Fref{Fig8}. When the tunneling ionization is the dominant mechanism,
in both atoms the mean tunneling time $\la\tau\ra\simeq0$.

\subsection{Classical photoelectron scattering}

A significant increase of the attoclock offset angle in the
low-intensity regime can be attributed to a thicker potential barrier,
as is illustrated in \Fref{Fig1}b). At the same time, a weaker field
drives the photoelectron to the detector slower, thus subjecting it to
a prolonged action of the Coulomb force. To disentangle these two
competing processes, \cn{2018PRLBray} devised a model in which the
classical scattering of the photoelectron in the Coulomb potential is
considered. The scattering angle in the attractive potential $V(r) =
-Z/r$ is given by the Rutherford formula \cite{Landau1982}.
\be
\tan{\theta\over2} = {1\over v_\infty^2}{Z\over \rho}
\ \ , \ \ 
\rho = {I_p\over E_0}
\ \ , \ \ 
v_\infty=A_0={E_0\over \w}
\ .
\label{Rutherford}
\ee
The distance of the closest approach $\rho$ in this expression is
equated in this model with the Keldysh tunnel width expressed via the
ionization potential $I_p$ and the maximum field strength $E_0$. The
photoelectron velocity at the detector $v_\infty$ is determined by the
peak value of the vector potential.  With these assumptions, the
attoclock offset angle in the case of the pure Coulomb potential takes
the form
\be
\theta_A=\frac12 \theta\simeq
{\w^2\over E_0^2}{Z\over \rho}
=
{\w^2\over E_0}{Z\over I_p} \ .
\label{KR}
\ee
In the above expression, the action of the laser field during the
photoelectron propagation to the detector is neglected. This
assumption is only valid for very short and weak laser pulses. The key
feature of the KR model is its intensity dependence $I^{-1/2}$ which
explains the growth of the attoclock angle $\theta_A$ in the low field
regime due to a greater elastic scattering of a slower photoelectron
in the Coulomb field. This field dependence is drawn in \Fref{Fig7}.

\begin{figure}[h!]

\hs{-0.cm}
\includegraphics[height=5.6cm]
                {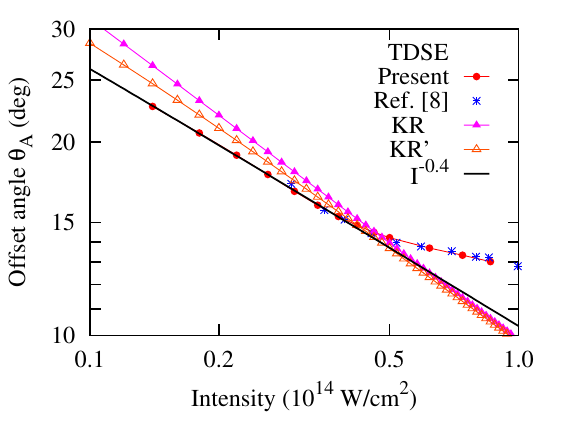}
\hs{-0cm}
\includegraphics[height=5.6cm]
                {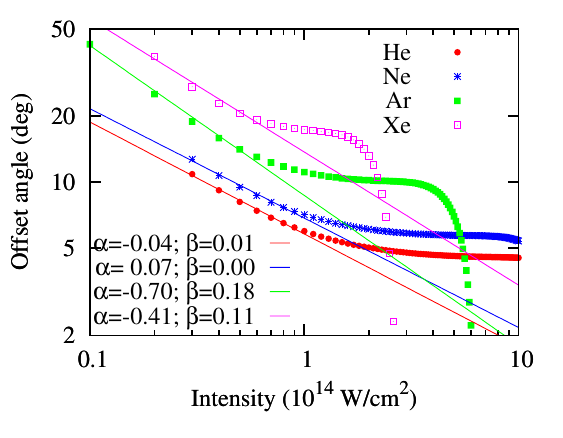}
\hs{-1.5cm}

\caption{Left: The attoclock offset angle $\theta_A$ of the hydrogen
  atom driven by a single-oscillation lather pulse. The results of the
  TDSE calculations of \cn{2018PRLBray} and \cn{2015NatPhysTorlina}
  are displayed. A comparison with the prediction of the KR model is
  made.
Right: the attoclock offset angles $\theta_A$ of noble gas atoms
driven by a single-oscillation lather pulse. The CTMC calculation by 
\cn{0953-4075-50-5-055602} is fitted with a generalized KR ansatz
 Adapted from \cn{2018PRLBray}
}
\label{Fig10}
\end{figure}

Thus defined Keldysh-Rutherford (KR) model was validated in the
numerical attoclock settings on hydrogen against numerical TDSE
calculations with a single oscillation laser pulse. Results of this
validation are shown in the left panel of \Fref{Fig10}. Here TDSE
calculations of \cn{2018PRLBray} and the H2 set of
\cn{2015NatPhysTorlina} are shown to be nearly indistinguishable.
This is compared with the two KR and KR$'$ estimates. The KR refers to
\Eref{KR}, whereas the KR$'$ does not make the small angle
approximation for the tangent function.  The KR scales at all
intensities as $I^{-0.5}$ by construction.  Fitting the KR$′$ in the
low intensity range yields $I^{-0.44}$. The TDSE results display a
similar dependency  $I^{-0.41}$  for the same region but then
flatten and deviate from both the KR and KR$′$. This is understandable
as the KR model is expected to work for weak fields only when the
field-driven trajectory is close to that involved in the field-free
scattering. Within the range of its validity, the KR model attributes
nearly all of the attoclock offset angle $\theta_A$ to the Coulomb
scattering. 

Although the KR model is developed explicitly for hydrogenic targets,
it can be also applied to other atoms. Indeed, the asymptotic charge
affecting the departing photoelectron is always the same $Z=1$ for all
neutral atomic targets. So the basic premise of the KR model remains
valid. To test this model for other atoms, an extended set of
numerical attoclock simulations \cite{0953-4075-50-5-055602} was
chosen. These simulations are conducted using the CTMC method.  In the
right panel of \Fref{Fig10}, the CTMC offset angles are fitted with
a generalized KR ansatz
\be
\theta_A(I) =  \frac{\w^2}{ I_p}
\frac{(1+\alpha)}{(I/2I_0)^{0.5+\beta}}
 \ ,
\label{fit}
\ee
where $I_0=$\Wcm{3.51}{16} is one atomic unit of field intensity.
Parameters $\alpha$ and $\beta$ indicate the deviation of the CTMC
calculation from the KR predictions. For small intensities, the scaling
of the offset angles with the field intensity is indeed close to
$I^{-0.5}$. There is strongest deviation from the KR prediction in Ar and
Xe where the fitting parameters are comparatively large.  At the same
time, these parameters are close to zero for the targets with the
larger ionization potentials, He and Ne.

\subsection{Strong field approximation}

While the KR model neglects entirely the driving of the photoelectron
by the laser pulse beyond the tunnel exit, the strong field
approximation (SFA) neglects completely the Coulomb field of the ion
remainder.  Under the latter assumption, the photoelectron motion can
be traced along a few dominant trajectories by the time integration of
the quasi-classical action. This integration can be carried over by
the steepest descent technique using the saddle point method (SPM)
(Milo\^sevi\'c {\em et~al} 2002, 2006)
\nocite{PhysRevLett.89.153001,0953-4075-39-14-R01}. This approach is
justified when the electron action accumulated along a
quasi-classical trajectory is large, $S\gg\hbar$. This is usually the
case in strong low-frequency laser fields. 

The ionization amplitude in the SFA is written as (Milo\^sevi\'c 
{\em et~al} 2002, 2006)
\ba
\label{SPM}
\hs{-2cm}
D(\k) &=& -i\sum_{s=1}^{N_{\rm SP}}
\left\{
2\pi i\over \E(t_s)\cdot [\k+\A(t_s)]
\right\}^{1/2}
\la\k+\A(t_s)|\r\cdot\E(t_s)|\psi_0\ra \
\exp[iS_{\k}(t_s)] 
\ ,
\ea
where 
$
S_\k(t) = \int^t dt'
\{
[\k+A(t')]^2/2+I_p
\}
$
is the semi-classical action.  The summation in \Eref{SPM} is carried
over $N_{\rm SP}$ saddle points $t_s$ that are solutions of the saddle
point equation
\be
\label{SP}
\partial S_{\k}(t_s)/\partial t
=
[\k+\A(t_s)]^2/2+
I_p =0
\ .
\ee
%
%
%
For circular polarization, the number of the saddle points $N_{\rm
  SP}=N+1$, where $N$ is the number of the pulse oscillations
\cite{0953-4075-39-14-R01}. With the presently chosen envelope
\eref{envelope} with $N=2$, $N_{\rm SP}=3$ of which only one dominant
SP makes the overwhelming contribution to the PMD shown on the bottom
right panel of \Fref{Fig3}.

\begin{figure}[h!]

\hs{1.cm}
\includegraphics[height=5.5cm]
                {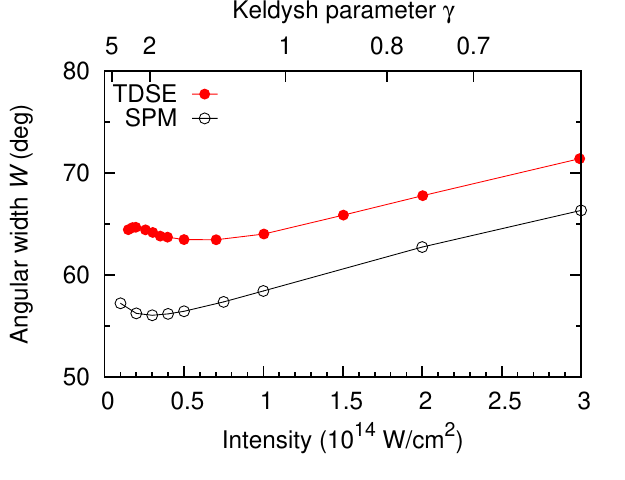}
\hs{1cm}
\includegraphics[height=5cm]
                {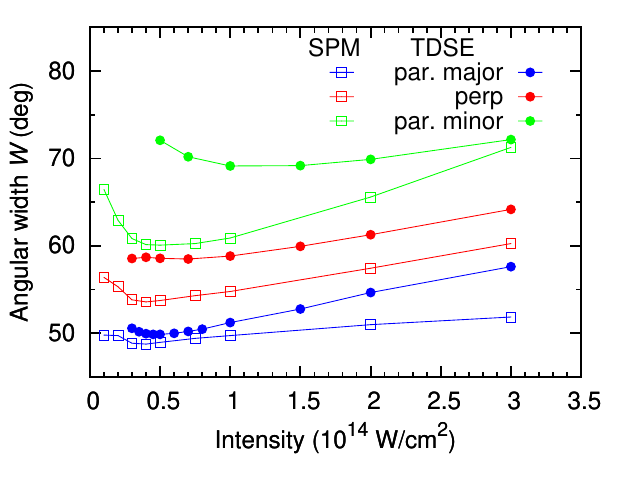}

\vs{-3mm}
\caption{Left: Angular width ${\cal W}(I)$ of the PMD in the
  polarization plane as a function of the field intensity $I$ for the
  atomic (left) and molecular (right) hydrogen.  The TDSE and PMD
  calculations are shown.  The \H molecule is aligned perpendicular to
  the polarization plane (red), parallel to the major polarization
  axis $\hat{\bm{e}}_y$ (black) and parallel to the minor polarization
  axis $\hat{\bm{e}}_x$ (green). Adapted from \cn{PhysRevA.99.063428}
\label{Fig11}}
\vs{-3mm}
\end{figure}

The main difference between the PMDs shown on the left and right
bottom panels of \Fref{Fig3} is that $\theta_A=0$ in the SFA with SPM.
This has long been a well-known fact, see
e.g. \cn{0953-4075-42-16-161001}. This fact serves as another
indication of the main contribution to the attoclock offset angle
coming from the Coulomb field of the residual ion. This field is
neglected in the SFA. Except for the vanishing offset angle, the
overall structure of the PMD in the polarization plane is reproduced
remarkably well by the SPM. This PMD can be quantified by its angular
width $\cal W$. This width is marked on the top panels of \Fref{Fig3}
between the fringes of the PMD ($f$-points) while the center of this
distribution is marked with the $c$-points. Numerically, the width
$\cal W$ is extracted from the Gaussian fitting to the radially
integrated momentum density. The width parameter ${\cal W}(I)$
extracted from the TDSE and SPM calculations are shown on the left
panel of \Fref{Fig11} as functions of the field intensity $I$.  This
dependence is not monotonous which can be qualitatively understood
from the SFA formulas given by \cn{Mur2001} for a continuous
elliptical field. In this case, the SP equation \eref{SP} can be
solved analytically.  For strong fields, when the Keldysh adiabaticity
parameter $\gamma\ll1$, the angular width grows with intensity. In the
opposite limit $\gamma\gg1$ the width is falling with intensity. The
minimum between these falling and rising intensity dependence of the
width occurs around $\gamma\simeq 1$.

The SPM technique can be easily adapted to the \H molecule. In this
case, the SPM equation \eref{SPM} acquires an additional term
$
(-1)^j\E(t_{sj})\cdot\R/2 \ .
$
This term defines the energy gain or loss for the electron to travel
to the molecule midpoint and has the opposite signs for different
atomic sites $j=1,2$. Accordingly, the single dominant SP in the
atomic case is split into two points. The corresponding factor
$
\exp[\pm i \k\cdot\R/2]
$
in the ionization amplitude defines the phase difference between the
two wave packets emitted from different atomic sites. The molecular
terms in both the SPM and the ionization amplitude vanishes when the
molecule is aligned perpendicular to the polarization plane while $\E$
and $\k$ are bound to this plane and hence the two solutions $t_{s1}$
and $t_{s2}$ become identical.

The effect of the molecular orientation is illustrated in the right
panel of \Fref{Fig11} where the angular width parameter $\cal W$ for
the \H molecule is displayed in three orientations: perpendicular to
the polarization plane (shown with red symbols), aligned with the peak
$\E_0$ field (``major'' axis, blue symbols) and with
the peak $\A_0$ potential (``minor'' axis, green
symbols).  Both the TDSE and SPM results are shown (filled circles and
open squares, respectively).
The angular width $\cal W$ varies very significantly depending on the
molecular orientation. Qualitatively, this behaviour is similar in the
TDSE and SPM calculations.  The latter model allows for the
understanding of this behavior qualitatively in terms of the
two-center interference.  For a given in-plane orientation the
interference term effectively increases (decreases) the ionization
potential thus increasing (or decreasing) the Im~$t_s$ and relative
contribution of the corresponding saddle points.

\section{Improved attoclock}
\markboth{Improved attoclock}{Improved attoclock}

When the tunneling ionization is induced by a long circularly
polarized laser pulse, the PMD in the polarization plane becomes a
circle. However, if a weak linearly polarized field is superimposed,
the maximum of the laser field is attained when the two fields are in
phase.  Because of an exponential sensitivity of the tunneling
ionization, such a superposition produces a  pronounced peak in
the PMD which can be easily traced, both experimentally and
numerically.

Such an ``improved attoclock'' experiment was conducted in the
laboratory and modeled theoretically by
\cn{PhysRevLett.123.073201}. They combined a circularly polarized
pulse of 25~fs at 800 nm and \Wcm{1.1}{14} with a linearly polarized
second harmonic at 400~nm and \Wcm{6}{12}.  The phase between the
two-color fields was precisely monitored. The photoelectron momentum
distribution of the argon atom was measured using the COLTRIMS
technique.  When the 800~nm field made a cycle, the two field vectors
overlapped once along the direction of the $z$ axis, and at this
instant, the electric field strength reached the maximum
$\E_0$. Accordingly, the $x$ axis, the direction of the maximum vector
potential $\A_0$, served as the attoclock hand.  The displacement of
the PMD peak relative to the attoclock hand was observed
experimentally and simulated numerically under several different
approximations.

In the crudest SFA simulation, the transition matrix element was
numerically integrated and the angle of the most probable momentum was
obtained strictly along the attoclock hand in the $-x$ direction, as
expected.
In a more advanced TDSE simulation with an effective potential
\be
V_{\rm eff} = −[1 +(Z-1) e^{-r/r_s}]r
\ ,
\ee
constructed to match the ionization potential of the Ar ($Z=18$) atom,
the most probable momentum pointed away from the attoclock hand
towards the experimentally observed maximum.

Furthermore, the SFA wave function $\psi$ was used to construct the
Wigner function
\be
W(v,r;t) = \pi^{-1}
\int dv' 
\psi^* (v+v',t)
e^{-2irv'}
\psi(v-v',t) 
\ . 
\ee
The latter was employed to determine the most probable momentum and
the time at the tunnel exit $x_{\rm exit}=I_p/E_0$. Thus defined exit
time and momentum were tested under the experimental conditions of
\cn{2017PRLCamus} and found to be very similar with the original
values. Having conducted this test, \cn{PhysRevLett.123.073201}
simulated their measurement and obtained a quite significant
longitudinal exit momentum of 0.4~au and an exit time over 200~as.
They used these values as initial conditions in their CTMC
simulations. The latter were conducted both with and without
considering the Coulomb potential in the Newtonian equation of
motion. When considering the Coulomb potential, the simulation using
the Wigner initial conditions reproduced the results of the experiment
and the TDSE. Without this potential, the PMD peaked strictly in the
attoclock hand direction without any visible offset.

These results confirm the conclusion that we have already reached in
this review several times. The offset angle of the attoclock, both in
its original or an improved configuration, is attributed wholly to the
effect of the Coulomb field of the ion reminder. The Wigner time, no
matter how large it could be, is rather meaningless. It has nothing to
do with the angular reading of the attoclock which is not, in fact, a
clock but rather a very precise ``nano-ruler''.

\section{Conclusion and outlook}
\markboth{Conclusion and outlook}{Conclusion and outlook}

We reviewed recent results related to attoclock measurements and
calculations on various atomic and molecular targets. Our main
emphasis was on the determination of the tunneling time, i.e. the time
the tunneling electron spends under the classically inaccessible
barrier. This interval of time is measured between the peak of the
electric field, when the bound electron starts tunneling, and the
instance the photoelectrin exits the tunnel.  At this instance, the
photoelectron momentum captures the vector potential of the driving
laser pulse. In the experiment, the tunneling time is extracted from
the offset angle between the angular maximum of the photoelectron
momentum distribution in the polarization plane and the attoclock
hand. This hand points along the vector potential direction $\A_0$ at
the instant of tunneling ionization when the electric field of the
laser pulse $\E_0$ is at its peak value. An alternative explanation of
this angular offset is due to the photoelectron scattering in the
Coulomb field of the ion remainder.  Various arguments are presented
here in support of the latter interpretation. Firstly, the offset
angle vanishes when the Coulomb potential is substituted with its
short-range Yukawa counterpart. The offset angle in the Yukawa atom
and the Yukawa molecule are as small as $0.25^\circ$
\cite{2019NatSainadh} and $2^\circ$ \cite{Wei2019},
respectively. Similarly, the offset angle is absent in negative ions
\cite{2019PRADouguet}.  Second, the offset angle is missing entirely
in the strong field approximation when the Coulomb field is neglected
\cite{0953-4075-42-16-161001,PhysRevA.99.063428}. Lastly, nearly all
of the rapid growth of the offset angle in the low laser field regime
is attributed to the Coulomb potential \cite{2018PRLBray}. A gap
between the naked Coulomb and a hard screened Yukawa potentials can be
spanned continuously by varying the screening length.

Several numerical attoclock simulations with a short, nearly
single-oscillation laser pulse, return specific estimates of the
tunneling time
\cite{2015NatPhysTorlina,PhysRevLett.117.023002,PhysRevA.97.013426}. In
the absence of the target atom depletion and well in the tunneling
ionization regime $\gamma<1$, this time is close to zero. The
depletion  effect may cause the effective tunneling time to be
negative, i.e. the tunneling process starts before the electric field
of the driving pulse reaches its maximum. When the electric field is
at its peak, there are no bound electrons left in the target atom to
tunnel. Meanwhile, the effective tunneling time is never positive,
i.e. there is no delay in the tunneling process which is therefore
instantaneous.

There are two experimental observations which could be interpreted in
terms of a finite tunneling time. The helium atom measurement in the
low field regime \cite{PhysRevLett.111.103003,2014OpticaLandsman}
returned very large offset angles which were at variance with the
semi-classical modeling assuming instantaneous tunneling. This
measurement was also in strong disagreement with fully {\i ab initio}
calculations \cite{PhysRevA.89.021402,Scrinzi2014} in which no
assumptions on the tunneling scenario was assumed. The relative Ar
versus Kr measurement by \cn{2017PRLCamus} returned an
intensity-dependent offset angle difference. This difference could
only be interpreted by assuming a finite ``Wigner'' tunneling time
\cite{PhysRevA.90.012116}. However, a closer analysis of the data
of  \cn{2017PRLCamus} conducted in this review indicates that most of the inter-atomic
offset angle difference come from the Coulomb potential
contribution. Indeed, the tunnel exit location, determined by the
ionization potential, is different in these target atoms. Hence the
attoclock is, in fact, a very fine ``nano-ruler'' which is extremely
sensitive to the width of the potential barrier.  The same conclusion
was reached in the ``improved attoclock'' setting by
\cn{PhysRevLett.123.073201}.

Because of its sensitivity to the tunnel width, an attoclock
measurement can be used as a useful probe of fine details of atomic
and molecular potentials.  The application of the attoclock technique
to molecular targets has already begun. Theoretical results on the \H
molecule show a strong sensitivity of the numerical attoclock to the
molecular orientation \cite{PhysRevA.99.063428}. The laboratory
attoclock reports on \H are in waiting \cite{Satya2018,Wei2019} and
will be presented soon.

\section*{Acknowledgment} 
\markboth{Acknowledgment}{Acknowledgment}

The author gratefully acknowledges Alex Bray for his help in
preparation of the manuscript and providing several graphical
illustrations.  The author has also benefited greatly from many
stimulating discussions with Igor Litvinyuk, Robert Sang, Satya
Sainadh, Igor Ivanov, Vladislav Serov, XiaoJun Liu and Wei Quan. The
author is thankful to Vladislav Serov for critical reading of the
manuscript . Serguei Patchkovskii is acknowledged for
making his TDSE code available to our group.  Resources of the
National Computational Infrastructure (NCI) Facility were utilized.

\np

\section{References}
\markboth{References}{References}

\end{document}